\documentclass[aps,prl,twocolumn,superscriptaddress,showpacs,amsmath,amssymb]{revtex4}
\usepackage{graphicx}
\usepackage{longtable}
\usepackage{bm}

\begin{document}

\title{Unconventional Landau Levels in Bulk Graphite Revealed by Raman Spectroscopy}

\author{A. F. Garc\'{i}a-Flores}
\affiliation{Institute of Physics ``Gleb Wataghin'', P.O. Box 6165, University of Campinas - UNICAMP, 13083-970, Campinas, SP, Brazil}

\author{H. Terashita}
\affiliation{Institute of Physics ``Gleb Wataghin'', P.O. Box 6165, University of Campinas - UNICAMP, 13083-970, Campinas, SP, Brazil}

\author{E. Granado}
\email{egranado@ifi.unicamp.br}
\affiliation{Institute of Physics ``Gleb Wataghin'', P.O. Box 6165, University of Campinas - UNICAMP, 13083-970, Campinas, SP, Brazil}

\author{Y. Kopelevich}
\affiliation{Institute of Physics ``Gleb Wataghin'', P.O. Box 6165, University of Campinas - UNICAMP, 13083-970, Campinas, SP, Brazil}

\begin{abstract}

The electronic Raman scattering of bulk graphite at zero magnetic field reveals a structureless
signal characteristic of a metal. For $T\lesssim 100$ K and $B > 2$ T,
several peaks at energies scaling linearly with magnetic field were observed and ascribed to transitions
from the lowest energy Landau level(s) (LL) to excited states belonging to the same ladder.
The LLs are equally (unequally) spaced for high (low) quantum numbers, being surprisingly
consistent with the LL sequence from massive Dirac Fermions ($m^* = 0.033(2) m_e$)
with  Berry's phase $2 \pi$ found in graphene bilayers.
These results provide spectroscopic evidence that much of the unconventional physics recently revealed
by graphene multilayers is also shared by bulk graphite.

\end{abstract}


\maketitle

Despite being exhaustively investigated since the early days of condensed matter physics, graphite
has attracted renewed interest since the recognition that much in its physics was missed in the past \cite{Kopelreview}.
Among the recently discovered phenomena are (i) the metal-insulator transition (MIT) driven by magnetic field applied
perpendicular to graphene planes \cite{KopelMIT1,Du,KopelMIT2}; (ii) quasi-two-dimensional Dirac fermions (DF) occupying
an unexpectedly large portion of the Fermi surface \cite{KopelDirac1}; and (iii) the coexistence of quantum Hall effects
originating from massless DF and massive quasi-particles \cite{KopelMIT1,KopelDirac2}.
The occurrence of massless DF is also characteristic of
graphene monolayers \cite{NovoselovDF}, indicating that much of the graphene physics may be found in graphite.
This conclusion has been reinforced by a number of recent spectroscopic observations, such as angle resolved photoemission spectroscopy
(ARPES) \cite{ARPES} and magnetotransmission \cite{Orlita} experiments, which showed coxisting massless Dirac holes and massive electrons at the $H$ and $K$ points of the Brilloin
zone, respectively. The striking qualitative correspondence between the MIT found in graphite \cite{KopelMIT1,Du,KopelMIT2}
and graphene \cite{Checkelsky} strongly suggests the same physical mechanism behind of the observed phenomenon in both systems. Also, scanning tunneling spectroscopy (STS) experiments showed a Landau Level (LL) spectrum
characteristic of coexisting massless and massive DF \cite{Li}. Since STS experiments probed the
LLs at the surface, direct spectroscopic observations of LLs from massive DF in bulk graphite are still missing.

For non-relativistic band electrons within the effective mass ($m^*$) approximation,
the energy levels are quantized upon the action of a magnetic field in a series of equally spaced LLs,
\begin{equation}
E_n-E_0=n \hbar \omega_{c}, \label{eq1}
\end{equation}
where $\omega_{c}=eB/m^*$ is the cyclotron frequency and $E_0=\frac{1}{2} \hbar \omega_{c}$ is the lowest energy level. In this work, we denote the LL ladders
satisfying Eq. \ref{eq1} as `conventional' LLs. On the other hand, for a single graphene layer, the energy dispersion
is linear with momentum, $E=v_F \hbar k$ (massless relativistic electrons), and the LLs follow a distinct $B-$ and $n-$dependence,
\begin{equation}
E_n-E_0= \sqrt{2e\hbar v_F^2 n B}. \label{eq2}
\end{equation}
Here $E_0=0$ is the lowest, zero energy level. Equations \ref{eq1} and \ref{eq2} are therefore associated with distinct
kinds of quantum Hall effect (QHE), with Berry's phase 0 and $\pi$, respectively. A third member of the small family of
QHE systems was recently found for graphene bilayers, with Berry's phase $2\pi$ and a LL structure given by \cite{McCann,NovoselovBilayer}
\begin{equation}
E_n-E_0= \hbar \omega_{c} \sqrt{n(n-1)}, \label{eq3}
\end{equation}
where $E_0=E_1=0$.

\begin{figure}
\includegraphics[width=0.45 \textwidth]{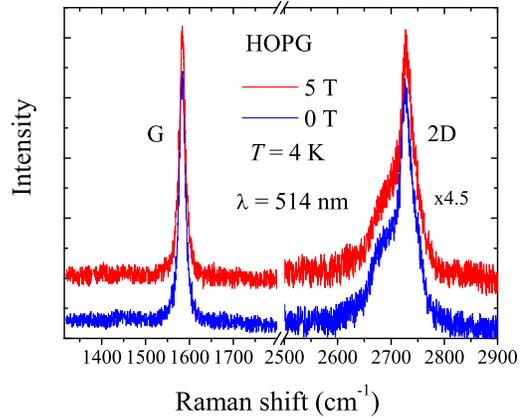}
\caption{\label{phonons} (Color online) Phonon Raman spectra of HOPG at 4 K and $B=0$ and $5$ T, showing its typical G mode and the 2D band. The spectrum at 5 T was vertically translated for clarity.}
\end{figure}

Raman scattering (RS) from LLs has been known both theoretically \cite{Wolff,Yafet,Mishchenko} and experimentally
\cite{Slusher,Patel,Worlock,Fainstein,Richards} for more than 40 years, and has been mostly observed
in semiconductor bulks and superstructures. In parallel, it is interesting to note that RS has been extensively used as a very
powerful and versatile characterization tool of graphitic materials \cite{Tuinstra,Thomsen,Saito,Reich,Gupta,Ferrari,Yan}, with a double
resonance mechanism \cite{Thomsen} for activation of a second-harmonic phonon band, so-called $2D$, connecting phononic and electronic structures. It is believed that RS measurements in graphite probes $\sim 20$ atomic layers \cite{Gupta}, being therefore bulk sensitive.
Surprisingly, no RS study of LLs in graphite have been performed, to the best of our knowledge.
In this paper, we present a temperature and magnetic field dependent RS study of graphite.
The analysis of the Raman data revealed a LL ladder following Eq. \ref{eq3}.
Our results provide new insights into the LL structure of bulk graphite and its relationship
with graphene bilayers.

The highly oriented pyrolitic graphite (HOPG) (from
the Research Institute ``Graphite'', Moscow) and Kish graphite samples studied in this work were thoroughly characterized by
means of x-ray, magnetotransport, magnetization, and ARPES \cite{KopelMIT1,KopelDirac1,KopelDirac2,ARPES}. 
Fresh surfaces were used, obtained by cleaving the samples in air.     
The RS spectra were excited with the 488 nm laser line from an argon-ion (Ar$^{+}$)
laser, unless otherwise noted, with an average power of $\sim 7$ mW focused in a spot of $\sim 100$ $\mu$m diameter.
The scattered
light was analyzed by a triple 1800 mm$^{-1}$ grating monochromator system in the subtractive mode
equipped with a N$_{2}$-cooled CCD detector. 
The studied samples were inserted within a commercial superconducting magnetocryostat with quartz windows. The magnetic field
($0 \leq B \leq 6.5$ T) was applied parallel to the hexagonal $c$-axis of graphite. All measurements were made in a near-backscattering configuration with the incident light propagating along the $c$ axis, and the scattered light being captured
within a solid angle of 0.15 sr.

\begin{figure}
\includegraphics[width=0.45 \textwidth]{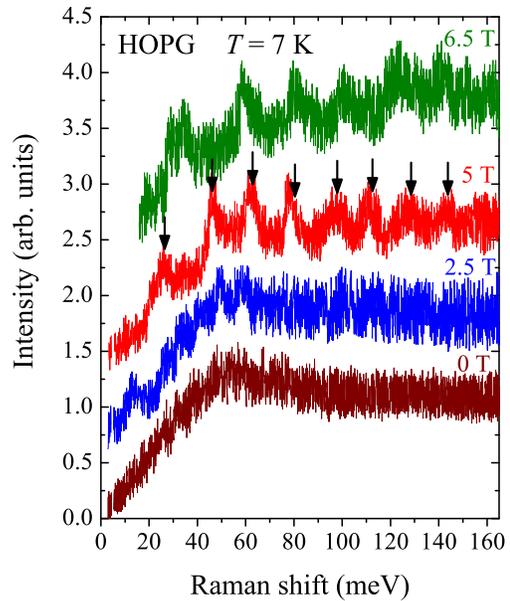}
\caption{\label{spectra} (Color online) Unpolarized electronic Raman spectra of HOPG at 7 K for selected applied magnetic fields. All Raman spectra, except at 0 T, were vertically translated for clarity. The observed peaks at 5 T are marked by arrows.
}
\end{figure}

Characteristic RS modes of bulk graphite were detected, as shown in Fig. \ref{phonons}. The absence of the forbidden $D$ peak at $\sim$1355 cm$^{-1}$ \cite{Tuinstra} within our sensitivity is indicative of a small defect concentration. Also, the Raman $G$ mode and $2D$ band did not present any significant
change with the applied magnetic fields in the studied range. This indicates
that the roughening of the electronic structure associated with the LL quantization caused by the field produces no observable effect
in the vibrational properties and in the double resonance mechanism \cite{Thomsen}, within our experimental resolution.

\begin{figure}
\includegraphics[width=0.45 \textwidth]{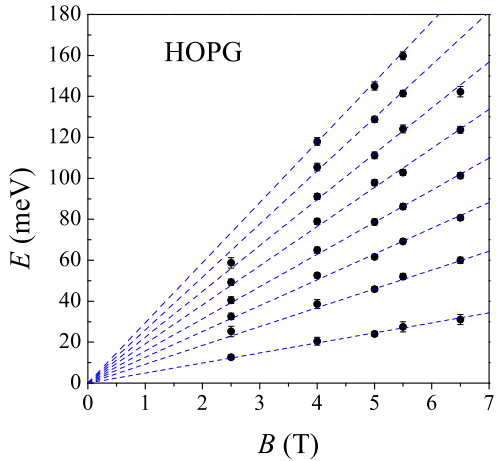}
\caption{\label{LLplot} (Color online) Peak positions obtained from the spectra of Fig. \ref{spectra} as a
function of magnetic field. Dotted lines represent linear fits crossing the origin.}
\end{figure}

The effect of the magnetic field can be most clearly seen
in the low-energy side of the RS spectra (see Fig. \ref{spectra}) At zero field, a structureless electronic continuous approaching null scattering as $\omega \rightarrow 0$ is observed,
which is characteristic of a normal metal \cite{Abrikosov}. This observation indicates that the relatively low
carrier density ($\sim 10^{18}$ cm$^{-3}$) does not prevent the observation of a significant electronic Raman signal in graphite.
Under application of a field, a roughening of the electronic RS can be observed, envolving into a clear sequence of
well defined peaks as the field is increased above $\sim 3$ T. The Raman shifts of these peaks are the same using the laser lines of 488 nm and 514 nm, excluding any possibility of fluorescence being the origin of this signal.
The peaks with higher energies appear to be
equally spaced, while the first couple of peaks seems to deviate from this pattern. A more quantitative analysis can
be made when the energies of all identified features in Fig. \ref{spectra} are plotted as a function of $B$
(see Fig. \ref{LLplot}). Here, a clear linear field-dependence of all identified peaks is readly noticed, allowing
for an unambiguous identification of these Raman peaks as due to transitions between LLs. Also in Fig. \ref{LLplot},
a nearly constant separation between the more energetic LLs can be readily verified, indicating that the observed
LL transitions refer to a unique ladder, i.e., from carriers belonging to the same electronic band. Again,
a clear deviation from the equal spacings rule can be noticed for the less energetic levels in Fig. \ref{LLplot}.
Preliminary polarized measurements (not shown) indicate that the linear polarization of the LL scattering is $90^{\circ}$ rotated with
respect to the incident light.

\begin{figure}
\includegraphics[width=0.45 \textwidth]{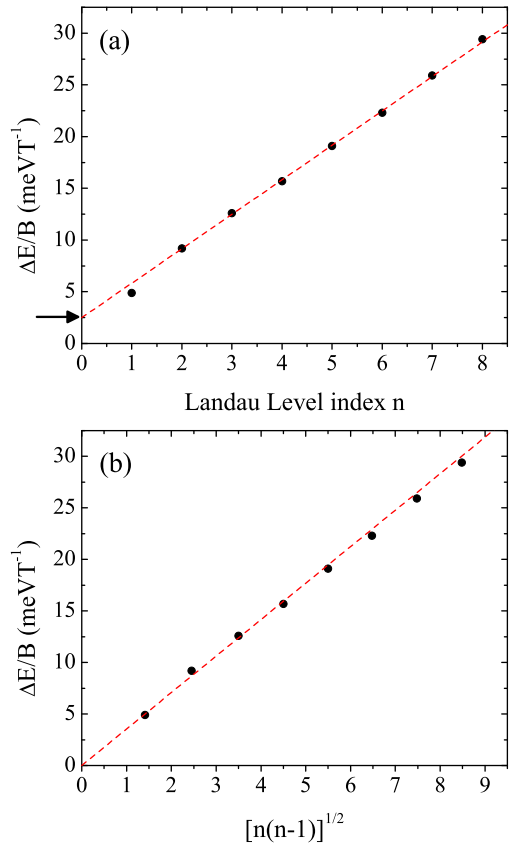}
\caption{\label{EBLL} (Color online) (a) $\Delta E/B \equiv (E_n - E_0)/B$ versus LL index $n$.
Solid lines are linear fits obtained for high $n$, which should cross the origin for conventional Landau levels
satisfying Eq. \ref{eq1} in the text. The arrow indicates this is not the case.
(b) $\Delta E/B$ versus $[n(n-1)]^{1/2}$ . The solid line represents the linear fit 
according to Eq. \ref{eq3} in the text.}
\end{figure}

Since the observed Raman peaks appear to belong to the same LL ladder, the indexing of the levels become straightforward.
Figure \ref{EBLL}(a) shows the slopes of the straight lines in Fig. \ref{LLplot} as a function of the level index $n$. Note that in a RS experiment, transitions between LLs are
observed, and therefore $E_n$ is not measured directly, but rather $\Delta E \equiv E_n - E_0$. For large $n$,
a linear behavior is observed. Nonetheless, the extrapolation of the straight line does not meet the origin.
This is a consequence of the deviation from the equal spacing law for small $n$, also evident in
Fig. \ref{EBLL}(a). Therefore the observed LL
sequence cannot be associated with a conventional LL ladder given by Eq. \ref{eq1}. On the other hand, in Fig. \ref{EBLL}(b), a
plot of $\Delta E/B$ versus $\sqrt{n(n-1)}$ yields a straight line crossing the origin, indicating that the observed LL
sequence is consistent with Eq. \ref{eq3}, yielding and effective mass $m^*=0.033(2) m_{e}$. Note that, in this case, the first observed level is indexed with $n=2$, since $E_0=E_1=0$ in Eq. \ref{eq3}. 

\begin{figure}
\includegraphics[width=0.45 \textwidth]{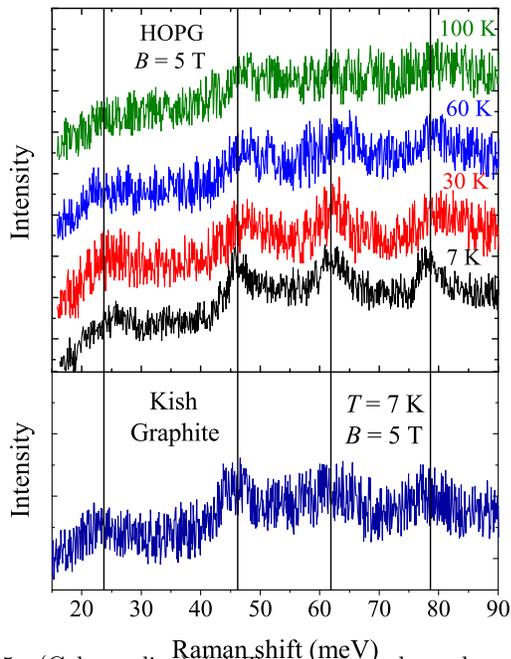}
\vspace{-1.2cm}
\caption{\label{Kish-HOPG} (Color online) (a) Temperature dependence of the Raman spectra of HOPG for $B$ = 5 T. (b) Raman spectra of  Kish graphite at 7 K and $B$ = 5 T.}
\end{figure}

The $T$- and sample-dependence of the LL spectra of graphite at 5 T were also studied (see Fig. \ref{Kish-HOPG}).
The LL peaks smear out uniformly for increasing $T$, being still observable at 100 K.
This is the highest $T$ at which LLs have been directly observed by a spectroscopic technique, to the best of our knowledge.
Also, a Kish graphite
sample was found to show the same LL spectrum as HOPG, within our resolution, indicating no considerable sample-dependence of our results.
We mention that STS results showed largely different surface LL spectra between Kish graphite
and HOPG \cite{Matsui}. 

Comparing our results to previous band structure calculations \cite{McClure,Dresselhaus,Nakao} and cyclotron resonance measurements \cite{Doezema}, the so-far observed bulk LLs of graphite do not appear to show a straightforward association with
our results. For instance, at the K-point and $B=5$ T, the separation between adjacent
levels is about 10 meV \cite{Nakao,Doezema}, while in our measurements this separation is between 20 meV (for high $n$) and 30 meV
(for low $n$) at this field. In this context,
it is interesting to notice that the LLs observed here and therefore the extracted
effective mass show a remarkable coincidence with a family of surface LLs in HOPG recently observed by STS measurements \cite{Li}, which
were identified as due to massive DF with Berry's phase $2\pi$. Our results lead to the unambiguous conclusion
that such unconventional LLs (and therefore the nature of the electrons causing such levels) are not restricted
to the surface, being actually bulk representative. The presence of massless DF with Berry's phase $\pi$ in bulk
graphite had been already demonstrated by de Haas - van Alphen and Shubnikov - de Haas oscilations \cite{KopelDirac1,KopelDirac2} and magnetotransmission studies \cite{Orlita}.
Our results yield spectroscopic evidence for unconventional LLs of a distinct nature in bulk graphite,
i.e., from massive DF with Berry's phase $2\pi$, complementing surface-sensitive STS measurements \cite{Li}.

It is intriguing that only unconventional LLs arising from massive DF were observed by RS.
In addition to such levels, others from massless DF \cite{KopelDirac1,KopelDirac2,ARPES,Orlita,Li} and also the classical
bulk LLs of graphite \cite{McClure,Dresselhaus,Nakao,Doezema} might have been observed. At this point, it is interesting to note
that previous observations of LLs from RS in semiconductors took place under resonance conditions \cite{Slusher,Patel,Worlock,Fainstein}. It is not implausible to infer that the LLs observed here were
enhanced by resonance in a semimetal such as graphite. Also, LLs arising from distinct carriers might show
largely different resonance enhancements of the scattering cross section, which in practice may have led to the observation
of a single ladder of LLs by RS. 

In summary, we observed unconventinal LLs in graphite by RS, consistent with the predicted spectra by
massive DF with Berry's phase $2\pi$. The fact that such carriers, characteristic of bilayer graphene,
as well as massless DF \cite{KopelDirac1,KopelDirac2,Orlita}, have been evidenced in bulk graphite deserves further theoretical analysis. For instance, Koshino and Ando \cite{Koshino} showed that the calculated LLs of graphene multilayers may be identical to monolayer or
bilayer graphene. It is therefore plausible that bulk graphite show coexisting LLs from both structures.

This work was supported by Fapesp and CNPq, Brazil.

\end{document}